\documentclass[preprint,showpacs,preprintnumbers,amsmath,amssymb]{revtex4}

\usepackage{graphicx}
\usepackage{dcolumn}
\usepackage{bm}
\usepackage{epsfig}

\begin{document}

\preprint{DESY~07-0667\hspace{12.55cm} ISSN 0418-9833}
\preprint{MZ--TH/07--07, LPSC 07--46\hspace{13.cm}}

\boldmath
\title{Finite-Mass Effects on Inclusive $B$-Meson Hadroproduction}
\unboldmath

\author{Bernd A. Kniehl}
\email{bernd.kniehl@desy.de}
\author{Gustav Kramer}
\email{gustav.kramer@desy.de}
\affiliation{{II.} Institut f\"ur Theoretische Physik,
Universit\"at Hamburg, Luruper Chaussee 149, 22761 Hamburg, Germany}
\author{Ingo Schienbein}
\email{schien@lpsc.in2p3.fr}
\affiliation{Laboratoire de Physique Subatomique et de Cosmologie,
Universit\'e Joseph Fourier Grenoble 1, CNRS/IN2P3, Institut National
Polytechnique de Grenoble, 53 avenue des Martyrs, 38026 Grenoble, France}
\author{Hubert Spiesberger}
\email{hspiesb@thep.physik.uni-mainz.de}
\affiliation{Institut f\"ur Physik, Johannes-Gutenberg-Universit\"at,
Staudinger Weg 7, 55099 Mainz, Germany}

\date{\today}

\begin{abstract}
We calculate the transverse-momentum ($p_T$) distribution for the inclusive
hadroproduction of $B$ mesons at intermediate values of $p_T$ at
next-to-leading order (NLO) in a dedicated finite-mass scheme using realistic
non-perturbative fragmentation functions that are obtained through a global
fit to $e^+e^-$ data from CERN LEP1 and SLAC SLC exploiting their universality
and scaling violations.
We find that finite-mass effects moderately enhance the cross section, by
about 20\% at $p_T=2m_b$, and rapidly fade out with increasing value of $p_T$,
so that the zero-mass prediction is reached.
We also perform comparisons with recent $p\bar{p}$ data taken by the CDF
Collaboration in run~II at the Fermilab Tevatron and comment on the usefulness
of the fixed-flavor-number scheme.
\end{abstract}

\maketitle

\section{Introduction}

Recently there has been much interest in the study of $B$-meson production in
$p\bar{p}$ collisions at hadron colliders, both experimentally and
theoretically.
The CDF Collaboration measured differential cross sections $d\sigma/dp_T$ for
the inclusive production of $B$ mesons (and their anti-particles) in
$p\bar{p}$ collisions at the Fermilab Tevatron as a function of the transverse
momentum $p_T$ in the central rapidity ($y$) region \cite{CDF,CDF1,CDF2,CDF3}.
The data reported in Ref.~\cite{CDF} were collected in the run period from
1992 to 1995 (runs IA and I) at a center-of-mass (c.m.) energy of
$\sqrt{S}=1.8$~TeV and were obtained using fully reconstructed $B^{\pm}$
mesons decaying into the exclusive final state $J/\psi K^{\pm}$.
The data presented in Ref.~\cite{CDF1} come from measurements in run II with
$\sqrt{S}=1.96$~TeV, where the inclusive differential production cross section
of $J/\psi$ mesons was used and the fraction of events from the decay of
long-lived $b$ hadrons was separated by analyzing the lifetime distribution.
These $b$ hadrons include $B^{+}$, $B^{-}$, $B^{0}$, and $\overline{B}^{0}$
mesons.
The data in Ref.~\cite{CDF2} were also taken at $\sqrt{S}=1.96$~TeV in run II.
In this case, the inclusive cross section for the production of $B^{\pm}$
mesons was obtained, as in Ref.~\cite{CDF}, by reconstructing
$B^{\pm}\to J/\psi K^{\pm}$ decays.
Very recently, CDF presented preliminary data from run~II based on events
with $B^-\to D^0\mu^-\overline{\nu}_\mu$ followed by $D^0\to K^-\pi^+$
and $B^-\to D^{*+}\mu^-\overline{\nu}_\mu$ followed by $D^{*+}\to D^0\pi^+$
and $D^0\to K^-\pi^+$ collected with the lepton-plus-displaced-track trigger
\cite{CDF3}.
These data explore the range 25~GeV${}<p_T<40$~GeV for the first time.
Although the measurement of $B$ mesons is experimentally well defined,
theoretical predictions did not agree with the data in the past.

In order to calculate the $B$-meson production cross section, the
non-perturbative fragmentation function (FF) for the transition $b\to B$ must
be known beforehand.
The QCD-improved parton model implemented in the modified minimal-subtraction
($\overline{\rm MS}$) renormalization and factorization scheme then provides a
rigorous theoretical framework for a coherent global data analysis.

In this framework, two distinct approaches for next-to-leading-order (NLO)
calculations in perturbative QCD have been used for comparisons with
experimental data.
In the so-called massless scheme or zero-mass variable-flavor-number scheme
(ZM-VFNS) \cite{BKK,BK,CG}, which is the conventional parton model approach,
the zero-mass-parton approximation is applied also to the $b$ quark, although
its mass $m$ is certainly much larger than the asymptotic scale parameter
$\Lambda_{\rm QCD}$.
In this approach, the $b$ quark is also treated as an incoming parton
originating from the (anti)proton, leading to additional contributions besides
those from $u$, $d$, $s$, and $c$ quarks and the gluon ($g$).
Although this approach can be used as soon as the factorization scales
associated with the initial- and final-state singularities are above the
starting scale of the parton distribution functions (PDFs) and the FFs, the
predictions are reliable only in the region of large $p_T$ values, with
$p_T\gg m$, where terms of the order of $m^2/p_T^2$ can safely be neglected.
A NLO calculation in this scheme automatically resums leading and
next-to-leading logarithms (NLL), {\it i.e.}\ terms of the form
$[\alpha_s\ln(p_T^2/m^2)]^n$ and $\alpha_s[\alpha_s\ln(p_T^2/m^2)]^n$ with
$n=1,2,3,\ldots$.
At the same time, all non-logarithmic terms through $\mathcal{O}(\alpha_s)$
relative to the Born approximation are retained for $m=0$ \footnote{%
It is, therefore, misleading to refer to NLO calculations in the ZM-VFNS as
NLL calculations, as is sometimes done.}.

The other calculational scheme is the so-called massive scheme or
fixed-flavor-number scheme (FFNS) \cite{theory}, in which the number of active
flavors in the initial state is limited to $n_f=4$, and the $b$ quark appears
only in the final state.
In this case, the $b$ quark is always treated as a heavy particle, not as a
parton.
The actual mass parameter $m$ is explicitly taken into account along with
$p_T$.
In this scheme, $m$ acts as a cutoff for the initial- and final-state
collinear singularities and sets the scale for the perturbative calculations.
A factorization of these would-be initial- and final-state collinear
singularities is not necessary, neither is the introduction of a FF for the
transition $b\to B$.
However, at NLO, terms proportional to $\alpha_s\ln(p_T^2/m^2)$, where
$\alpha_s$ is the strong-coupling constant, arise from collinear gluon
emissions by $b$ quarks or from branchings of gluons into collinear
$b\overline{b}$ pairs.
These terms are of order $O(1)$ for large values of $p_T$, and with the choice
$\mu_R={\cal O}(p_T)$ for the renormalization scale they spoil the convergence
of the perturbation series.
The FFNS with $n_f=4$ should thus be limited to a rather small range of $p_T$,
from $p_T=0$ to $p_T\agt m$.
The advantage of this scheme is that the $m^2/p_T^2$ power terms are fully
taken into account.

The ZM-VFNS and FFNS are valid in complementary regions of $p_T$, and it is
desirable to combine them in a unified approach that incorporates the virtues
of both schemes, {\it i.e.}\ to resum the large logarithms, retain the full
finite-$m$ effects, and preserve the universality of the FFs.
This is necessary for a reliable and meaningful interpretation of the CDF data
\cite{CDF,CDF1,CDF2}, which mostly lie in the transition region of the two
schemes.
An earlier approach to implement such an interpolation is the so-called
fixed-order-next-to-leading-logarithm (FONLL) scheme, in which the
conventional cross section in the FFNS is linearly combined with a suitably
modified cross section in the ZM-VFNS with perturbative FFs, using a
$p_T$-dependent weight function \cite{CGN,CN}.
Then the FONLL cross section is convoluted with a non-perturbative FF for the
$b\to B$ transition.
These FFs are adjusted to $e^+e^-$ data, using the same approach, and good
agreement with the CDF data was obtained.

In this work, we wish to present the results of an approach that is much
closer in spirit to the ZM-VFNS, but keeps all $m^2/p_T^2$ power terms in the
hard-scattering cross sections.
This scheme is called general-mass variable-flavor-number scheme (GM-VFNS) and
has recently been worked out for the photoproduction \cite{KS,KS1} and
hadroproduction \cite{KKSS,KKSS1,KKSS2} of charmed hadrons.
In this approach, one starts from the region $p_T\gg m$ and absorbs the large
logarithms $\ln(\mu_F^2/m^2)$, where $\mu_F$ is the factorization scale of the
initial or final state, into the $b$-quark PDF of the incoming hadrons and the
FF for the $b\to B$ transition.
After factorizing the $\ln m^2$ terms, the cross section is infrared safe in
the limit $m\to 0$, and $n_f=5$ is taken in the strong-coupling constant and
the Dokshitzer-Gribov-Lipatov-Altarelli-Parisi (DGLAP) evolution equations.
The remaining $m$-dependent contributions, {\it i.e.}\ the $m^2/p_T^2$ power
terms, are retained in the hard-scattering cross sections.
These terms are very important in the region of intermediate $p_T$ values,
$p_T\agt m$, and are expected to improve the theoretical predictions as
compared to the ZM-VFNS.
The large logarithms are absorbed into the PDFs and FFs by subtraction of the
collinearly (mass) singular terms at the initial- and final-state
factorization scales, respectively.

It is well known that the subtraction of just the collinearly, {\it i.e.}\
mass singular terms, does not define a unique factorization prescription.
Also finite terms must be specified.
In the conventional ZM-VFNS calculation, one puts $m=0$ from the beginning,
and the collinearly divergent terms are defined with the help of dimensional
regularization.
This fixes the finite terms in a specific way, and their form is inherent to
the chosen regularization procedure.
If one starts with $m\neq0$ and performs the limit $m\to0$ afterwards, the
finite terms are different.
These terms have to be removed by subtraction together with the $\ln m^2$
terms in such a way that, in the limit $p_T\to\infty$, the known massless
$\overline{\rm MS}$ expressions are recovered.
This matching procedure is needed, since we use PDFs and FFs defined in the
ZM-VFNS.
A subtraction scheme defined in this way is the correct extension of the
conventional ZM-VFNS to include $b$-quark (or similarly $c$-quark) mass effects
in a consistent way.
We actually include the $c$-quark contribution in the massless approximation,
{\it i.e.}\ we treat the $c$ quark as one of the light partons.
  
The results of our earlier work on charmed-hadron inclusive production by
$p\bar{p}$ scattering at NLO in the GM-VFNS \cite{KKSS,KKSS1,KKSS2} directly
carry over to $b$ hadrons.
Then, the $b$ quark is the heavy one, with mass $m$, while the $c$ quark
belongs to the group of light quarks, collectively denoted by $q=u,d,s,c$ in
the following, whose mass is put to zero.
Furthermore, we need PDFs and FFs implemented with $n_f=5$ in the
$\overline{\rm MS}$ factorization scheme.
Non-perturbative FFs for the transitions $a\to B^\pm$, where
$a=g,q,\bar q,b,\bar b$, were extracted at leading order (LO) and NLO already
several years ago \cite{BKK} using data for the scaled-energy ($x$)
distribution $d\sigma/dx$ of $e^+e^- \to B+X$ at $\sqrt{S}=91.2$~GeV measured
by the OPAL Collaboration at CERN LEP1 \cite{Opal}.

We note that our implementation of the GM-VFNS is similar to the
Aivazis-Collins-Olness-Tung (ACOT) \cite{ACOT} scheme formulated for the
initial state of fully inclusive deep-inelastic scattering.
The extension of this scheme to the inclusive production of heavy partons was
considered in Ref.~\cite{OST}, where the resummation of the final-state
collinear logarithms was only performed to LO and parton-to-hadron FFs were
not included.
A discussion of the differences between our approach and the one in
Ref.~\cite{OST} concerning the collinear subtraction terms can be found in
Ref.~\cite{KKSS1}.

This paper is organized as follows.
In Sec.~\ref{sec:two}, we introduce new NLO sets of $B$-meson FFs.
In Sec.~\ref{sec:three}, we numerically analyze the GM-VFNS predictions with
regard to the impact of the $m$-dependent terms and the relative importance of
the various partonic initial states.
In Sec.~\ref{sec:four}, we compare the predictions of the GM-VFNS, and also
those of the ZM-VFNS and FFNS, with CDF data from run~II
\cite{CDF1,CDF2,CDF3}.
Our conclusions are contained in Sec.~\ref{sec:five}.

\boldmath
\section{Non-perturbative $B$-Meson Fragmentation Function}
\unboldmath
\label{sec:two}

As input for the calculation of inclusive $B$-meson production cross sections
one needs a realistic non-perturbative FF describing the transition of the $b$
($\bar{b}$) quark into a $B$ meson.
Such a FF can be obtained only from experiment.
In Ref.~\cite{BKK}, LEP1 data for the distribution in the scaled $B$-meson
energy, $x=2E_B/\sqrt{S}$, from OPAL \cite{Opal} were fitted at LO and NLO in
the ZM-VFNS using three different ansaetze for the $b\to B$ FF at the starting
scale $\mu_F=\mu _0$ of the DGLAP evolution, including the ansatz by Peterson
{\it et al.}\ \cite{Peterson},
\begin{equation}
D(x,\mu_0^2)=N\frac{x(1-x)^2}{[(1-x)^2+\epsilon x]^2},
\label{eq:peterson}
\end{equation}
and the simple power ansatz \cite{Kartvelishvili:1985ac},
\begin{equation}
D(x,\mu_0^2)=Nx^\alpha(1-x)^\beta.
\label{eq:power}
\end{equation}
The best fit was obtained for the Peterson ansatz.
In Ref.~\cite{BKK}, the starting scale was taken to be $\mu_0 = 2m$, with
$m=5.0$~GeV.   
The $a\to B$ FFs for $a=g,q,\bar q$ were assumed to be zero at $\mu_F=\mu_0$
and generated through the DGLAP evolution to larger values of $\mu_F$.

In the meantime, new and more precise measurements of the cross section of
inclusive $B$-meson production in $e^+e^-$ annihilation on the $Z$-boson
resonance have been published by the ALEPH \cite{Aleph}, OPAL \cite{Opal1},
and SLD \cite{SLD} collaborations, which motivates us to update the analysis
of Ref.~\cite{BKK}.
This also gives us the opportunity to adjust some of the choices made in
Ref.~\cite{BKK}, to conform with the conventions underlying modern PDF sets.
In fact, for our numerical analysis, we use the NLO proton PDF set CTEQ6.1M,
based on the $\overline{\rm MS}$ prescription, by the Coordinated
Theoretical-Experimental Project on QCD (CTEQ) \cite{CTEQ}.
In this set, the $b$-quark PDF has its starting scale at $\mu_0=m$ with
$m=4.5$~GeV.
The mass values used in PDFs and FFs have, of course, to be chosen
consistently in order to avoid the appearance of terms proportional to
$\ln(\mu_0^2/m^2)$ in the NLO corrections.
While a shift in the starting scale from $\mu_0=2m$ with $m=5.0$~GeV to
$\mu_0=m$ with $m=4.5$~GeV changes the $b$-quark FFs only marginally at
$\mu_F$ values relevant for the $e^+e^-$ annihilation cross sections to be
used in the fit, it does have a significant effect on the $g\to B$ FF, which
greatly affects the cross section predictions for the Tevatron.
The size of the analogous effect for $D^{*+}$ FFs is investigated in Fig.~1 of
Ref.~\cite{Kniehl:2006mw}, which uses $\mu_0=m_c$, through comparison with
Ref.~\cite{Binnewies:1997xq}, which uses $\mu_0=2m_c$.
We thus perform a combined fit to these three data sets \cite{Aleph,Opal1,SLD}
using $\mu_0=m$ with $m=4.5$~GeV as in Ref.~\cite{CTEQ}.
Furthermore, we adopt from Ref.~\cite{CTEQ} the NLO value
$\Lambda_{\overline{\rm MS}}^{(5)} = 227$~MeV appropriate for $n_f=5$, which
corresponds to $\alpha_s^{(5)}(m_Z)=0.1181$.
As in Ref.~\cite{BKK}, we chose the renormalization and factorization scales
to be $\mu_R=\mu_F=\sqrt{S}$.
We use the ansaetze of Eqs.~(\ref{eq:peterson}) and (\ref{eq:power}) for the
$b\to B$ FF at $\mu_F=\mu_0$, while the $g,q\to B$ FFs are taken to vanish at
$\mu_F=\mu_0$ and are generated through the DGLAP evolution.
In order to obtain acceptable fits, we have to omit some of the data points in
the small-$x$ region.
Specifically, we only include the ALEPH data with $x\ge0.25$, the OPAL data
with $x\ge0.325$, and the SLD data with $x\ge0.28$.
At the other end of the $x$ range, we include all data points up to $x=1$.
Altogether we use 18, 15, and 18 data points of the ALEPH, OPAL, and SLD sets,
respectively.
Since we only include in the fit data from the $Z$-boson resonance, finite-$m$
effects, being of relative order $m_b^2/m_Z^2=0.2\%$, are greatly suppressed,
so that we are comfortably within the asymptotic regime where the GM-VFNS is
equivalent to the ZM-VFNS.

\begin{figure}[!thb]
\begin{center}
\epsfig{file=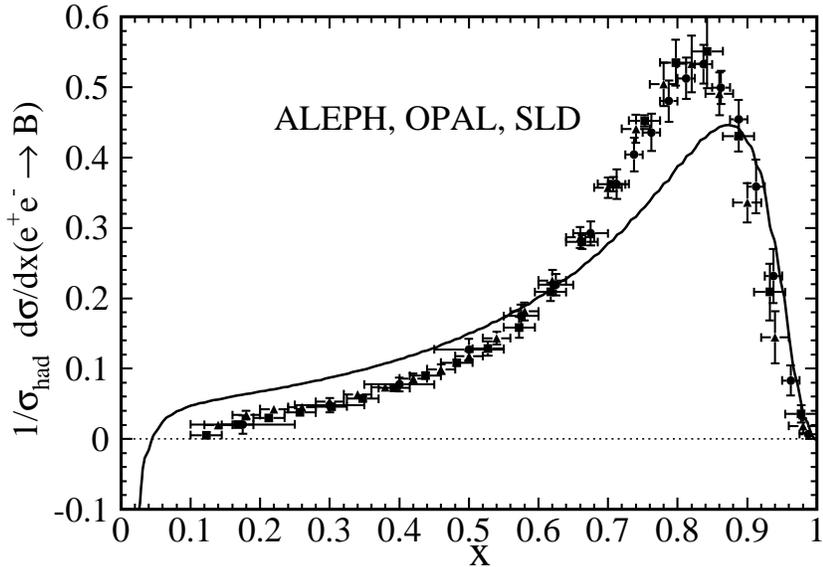,width=0.75\textwidth}
\vspace{-0.75cm}\\
(a)\\
\epsfig{file=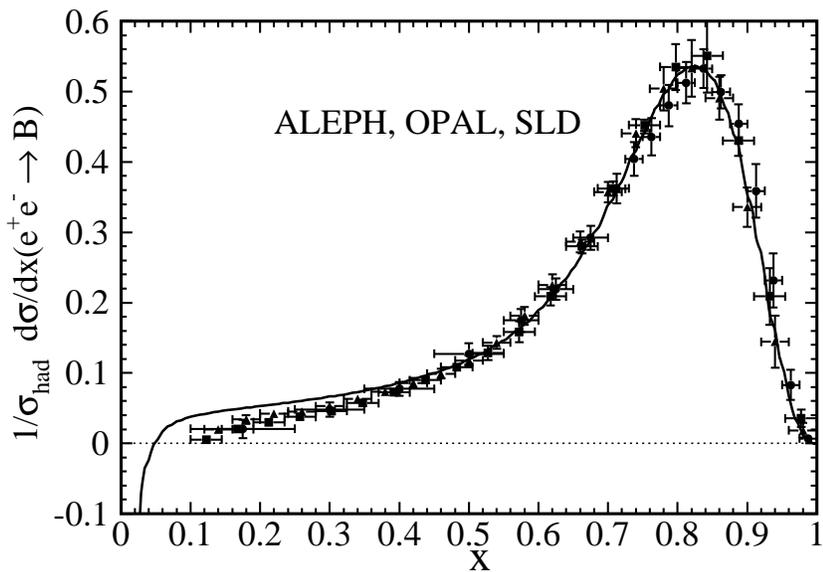,width=0.75\textwidth}
\vspace{-0.75cm}\\
(b)
\caption{Comparisons of the ALEPH \cite{Aleph} (circles), OPAL \cite{Opal1}
(squares), and SLD \cite{SLD} (triangles) data with the NLO fits using (a) the
Peterson ansatz~(\ref{eq:peterson}) and (b) the power ansatz~(\ref{eq:power}).
The initial factorization scale for all partons is $\mu_0=m=4.5$~GeV.
The different symbols can be better distinguished in the electronic edition,
where the figures can be enlarged.}
\label{fig1}
\end{center}
\end{figure}

\begin{table}
\begin{center}
\caption{Fit parameters of the $b$-quark FFs in Eqs.~(\ref{eq:peterson}) and
(\ref{eq:power}) at the starting scale $\mu_0=m=4.5$~GeV and values of
$\chi_{\rm d.o.f}^2$ achieved.
All other FFs are taken to be zero at $\mu_0=m$.}
\label{tab1}
\begin{ruledtabular}
\begin{tabular}{ccccc}
$N$ & $\alpha$ & $\beta$ & $\epsilon$ & $\chi^2/\mathrm{d.o.f}$ \\
\hline
0.06634 & -- & -- & 0.008548 & 21.37 \\
4684.1  & 16.87    & 2.628    & -- & 1.495 \\
\end{tabular}
\end{ruledtabular}
\end{center}
\end{table}
The values of the parameters in Eqs.~(\ref{eq:peterson}) and (\ref{eq:power})
obtained through the fits based on the Peterson and power ansaetze are listed
in Table~\ref{tab1} together with the respective values of $\chi^2$ per degree
of freedom, $\chi^2/\mathrm{d.o.f}$. 
The corresponding $d\sigma/dx$ distributions are compared with the ALEPH
\cite{Aleph}, OPAL \cite{Opal1}, and SLD \cite{SLD} data in Fig.~\ref{fig1}.
These three data sets mostly overlap and can hardly be distinguished in the
figure.
We observe from Table~\ref{tab1} and Fig.~\ref{fig1} that the power ansatz
yields an excellent overall fit to the selected data points.
There are deviations at $x\alt0.3$, which are due to the exclusion of data
points from the fit.
The $\chi^2/\mathrm{d.o.f}$ value for the combined fit is $1.495$.
The individual $\chi^2/\mathrm{d.o.f}$ values of the ALEPH, OPAL, and SLD data
sets are 0.861, 2.350, and 1.410, respectively.
On the other hand, the Peterson ansatz leads to an intolerable description of
the data, yielding $\chi^2/\mathrm{d.o.f}=21.37$ for the combined fit and
similar values for the individual data sets.
This ansatz has only two free parameters, $N$ and $\epsilon$, and is just not
flexible enough to account for the very precise experimental data.

\begin{table}
\begin{center}
\caption{Branching fractions $B(\mu_F)$ and average energy fractions
$\langle x\rangle(\mu_F)$ evaluated at factorization scales $\mu_F=4.5$, 9.0,
and 91.2~GeV using the $b\to B$ FFs based on the Peterson and power ansaetze.}
\label{tab2}
\begin{ruledtabular}
\begin{tabular}{ccccc}
 & \multicolumn{2}{c}{Peterson} & \multicolumn{2}{c}{power} \\
$\mu_F$ (in GeV) & $B(\mu_F)$ & $\langle x\rangle(\mu_F)$ & $B(\mu_F)$ &
$\langle x\rangle(\mu_F)$ \\
\hline
4.5 & 0.3994 & 0.8098 & 0.4007 & 0.8312 \\
9.0 & 0.3935 & 0.7542 & 0.3955 & 0.7730 \\
91.2 & 0.3767 & 0.6403 & 0.3803 & 0.6537 \\
\end{tabular}
\end{ruledtabular}
\end{center}
\end{table}
Besides the $b\to B$ FF itself, also its first two moments are of
phenomenological interest and subject to experimental determination.
They correspond to the $b\to B$ branching fraction,
\begin{equation}
B(\mu_F)=\int_{x_{\rm cut}}^1dx\,D(x,\mu_F^2),
\label{eq:bq}
\end{equation}
and the average energy fraction that the $B$ meson receives from the $b$
quark,
\begin{equation}
\langle x\rangle(\mu_F)=\frac{1}{B(\mu_F)}\int_{x_{\rm cut}}^1dx\,
xD(x,\mu_F^2),
\label{eq:xq}
\end{equation}
where the cut $x_{\rm cut}=0.15$ excludes the $x$ range where our formalism is
not valid.
We observe from Table~\ref{tab2} that the Peterson and power ansaetze lead to
rather similar results for $B(\mu_F)$ and $\langle x\rangle(\mu_F)$, the ones
for the power ansatz being slightly larger.
While $B(\mu_F)$ is practically independent of $\mu_F$,
$\langle x\rangle(\mu_F)$ is shifted towards smaller values through the
evolution in $\mu_F$.
It is interesting to compare the results for
$\langle x\rangle(91.2~\mathrm{GeV})$ in Table~\ref{tab2} with the values
quoted by ALEPH, OPAL, and SLD, which read
$0.7361\pm0.0061~(\mathrm{stat})\pm0.0056~(\mathrm{syst})$ \cite{Aleph},
$0.7193\pm0.0016~(\mathrm{stat})
\genfrac{}{}{0pt}{}{+0.0036}{-0.0031}~(\mathrm{syst})$
\cite{Opal1}, and
$0.709\pm0.003~(\mathrm{stat})\pm0.003~(\mathrm{syst})\pm
0.002~(\mathrm{model})$ \cite{SLD}, respectively.
We observe that the experimental results lie systematically above ours.
However, one must keep in mind that the experimental results refer to the
first moment of the measured cross section distribution $d\sigma/dx$, which
naturally includes all orders and also contributions from gluon and
light-quark fragmentation, while ours are evaluated from the $b\to B$ FF at
NLO in the $\overline{\rm MS}$ scheme via Eq.~(\ref{eq:xq}).
Of course, the $b\to B$ FF and its moments depend on scheme, order, and
implementation issues such as the functional form of the ansatz at the
starting scale $\mu_0$ and the value of $\mu_0$ itself, and thus do not
represent physical observables by themselves.
Nevertheless, comparisons of the quantities $B(\mu_F)$ and
$\langle x\rangle(\mu_F)$ defined in Eqs.~(\ref{eq:bq}) and (\ref{eq:xq}),
respectively, with their experimental counterparts are useful to check the
dominance of $b\to B$ fragmentation and are routinely performed in the
literature (see, {\it e.g.}, Ref.~\cite{BKK}).

\boldmath
\section{Theoretical Predictions for $p\bar{p} \rightarrow B+X$}
\unboldmath
\label{sec:three}

We are now in a position to perform a numerical analysis.
We consider the inclusive cross section of $p\bar{p} \rightarrow B + X$, where
$B$ stands for the average of the $B^+$ and $B^-$ mesons, at
$\sqrt{S}=1.96$~TeV as in run~II at the Tevatron.
We concentrate on the $p_T$ distribution integrated over $|y|<1$ corresponding
to the central region of the CDF detector.
We use the CTEQ6.1M proton PDFs \cite{CTEQ} and the $B$-meson FFs based on the
power ansatz presented in Sec.~\ref{sec:two}, both implemented at NLO with
$\Lambda^{(5)}_{\overline{\rm MS}} = 227$~MeV and $m=4.5$ GeV.
For simplicity, we use a common factorization scale for the initial and final
states. 
We set the renormalization and factorization scales to
$\mu_R=\xi_R m_T$ and $\mu_F=\xi_F m_T$, where $m_T =\sqrt{p_T^2+m^2}$ is the
transverse mass of the $b$ quark and $\xi_R$ and $\xi_F$ are introduced to
estimate the theoretical uncertainty.
Unless otherwise stated, we use the default values $\xi_R=\xi_F=1$.
With our default choices $\mu_0=m$ and $\mu_F=m_T$, we have $\mu_F\to\mu_0$ as
$p_T\to 0$.
In this limit, the FFs and $b$-quark PDF should fade out and quench the cross
section, leading to a turn-over of the $p_T$ distribution.
However, the precise location of the maximum and other details of the line
shape are also subject to other implementation issues of the GM-VFNS.
We shall return to this topic in Sec.~\ref{sec:four}.

\begin{figure}[!b]
\begin{center}
\begin{tabular}{ll}
\parbox{0.5\textwidth}{
\epsfig{file=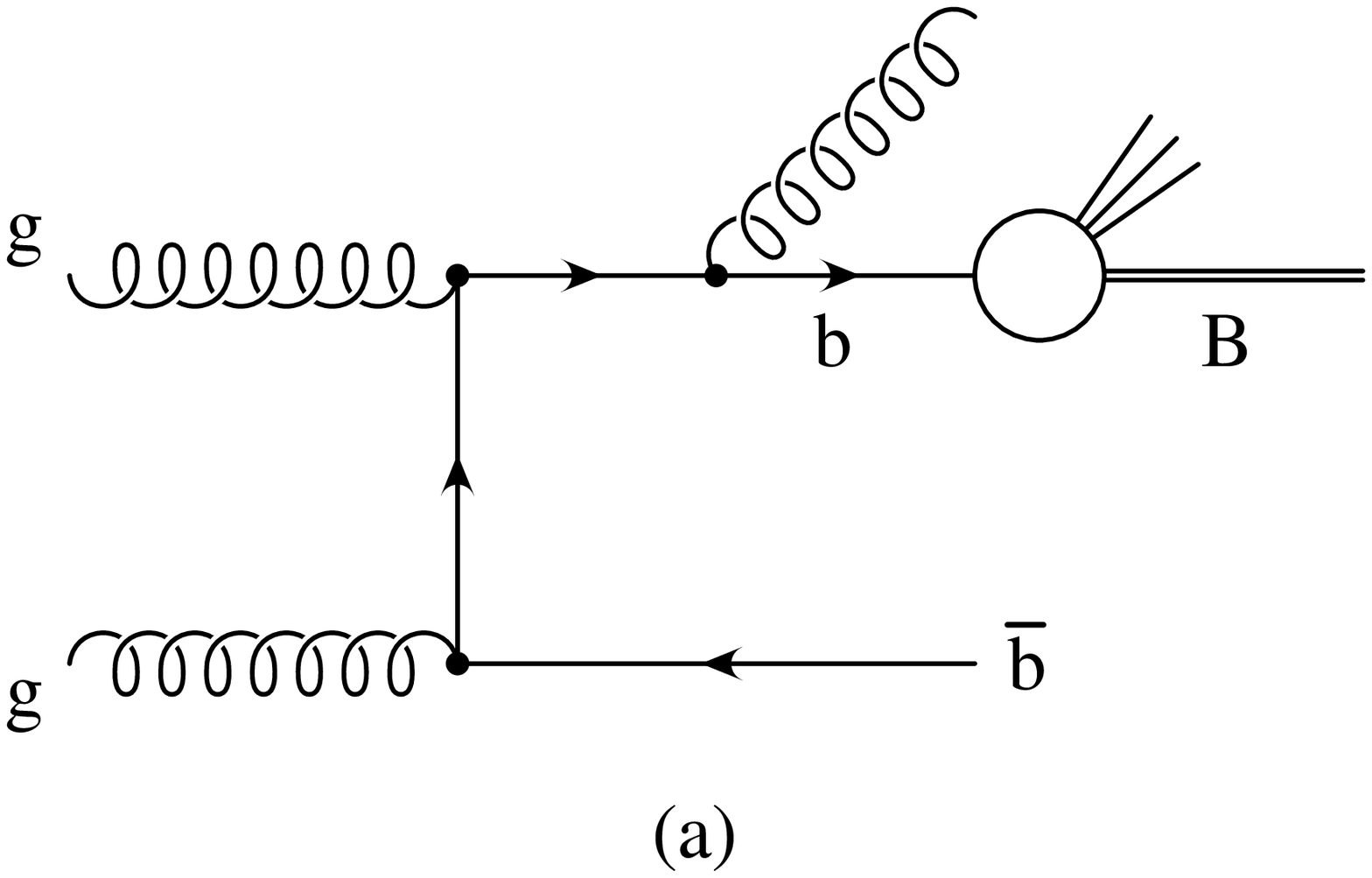,width=0.45\textwidth}}
&
\parbox{0.5\textwidth}{
\epsfig{file=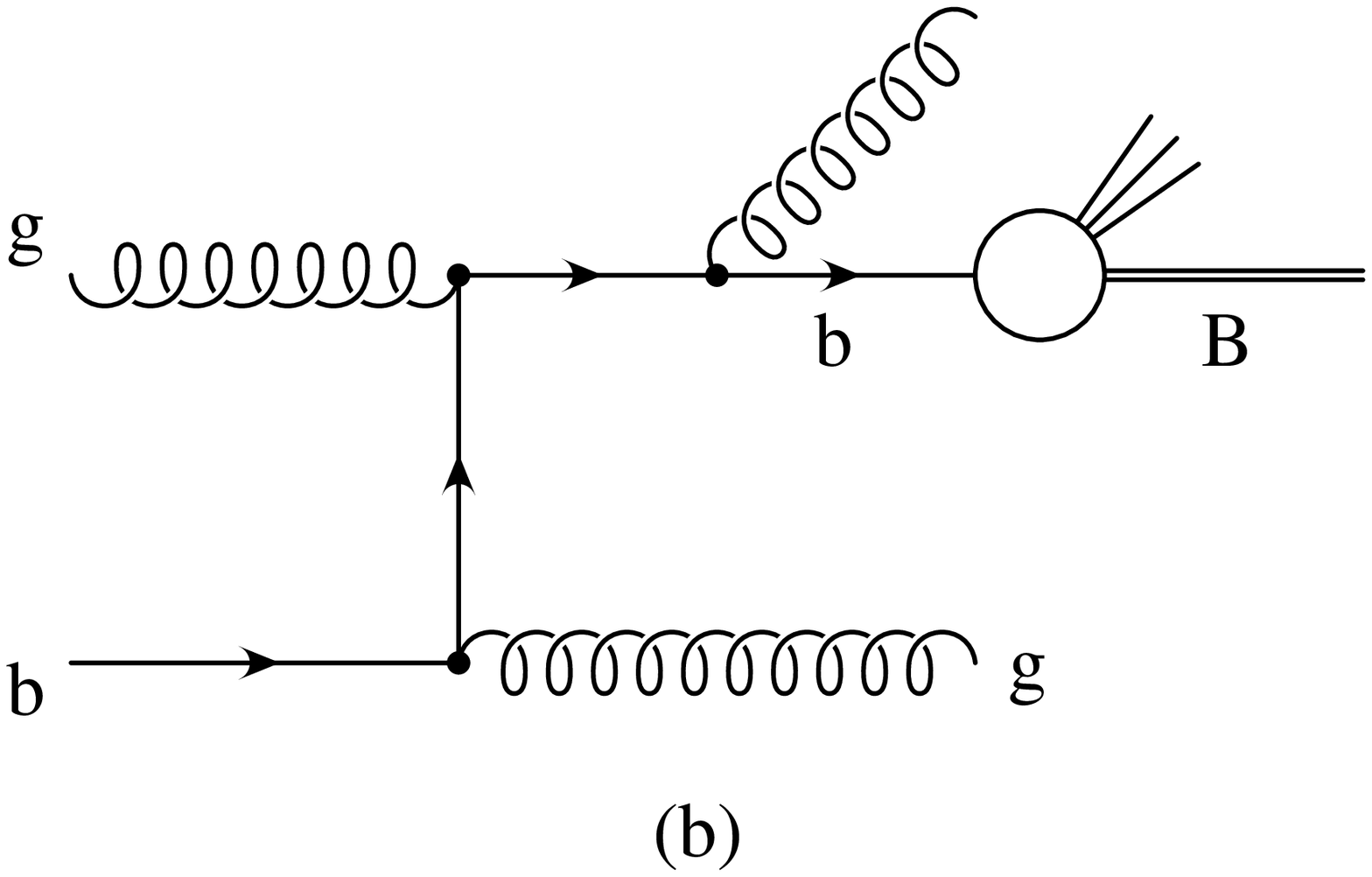,width=0.45\textwidth}
}
\end{tabular}
\begin{center}
\parbox{0.5\textwidth}{
\epsfig{file=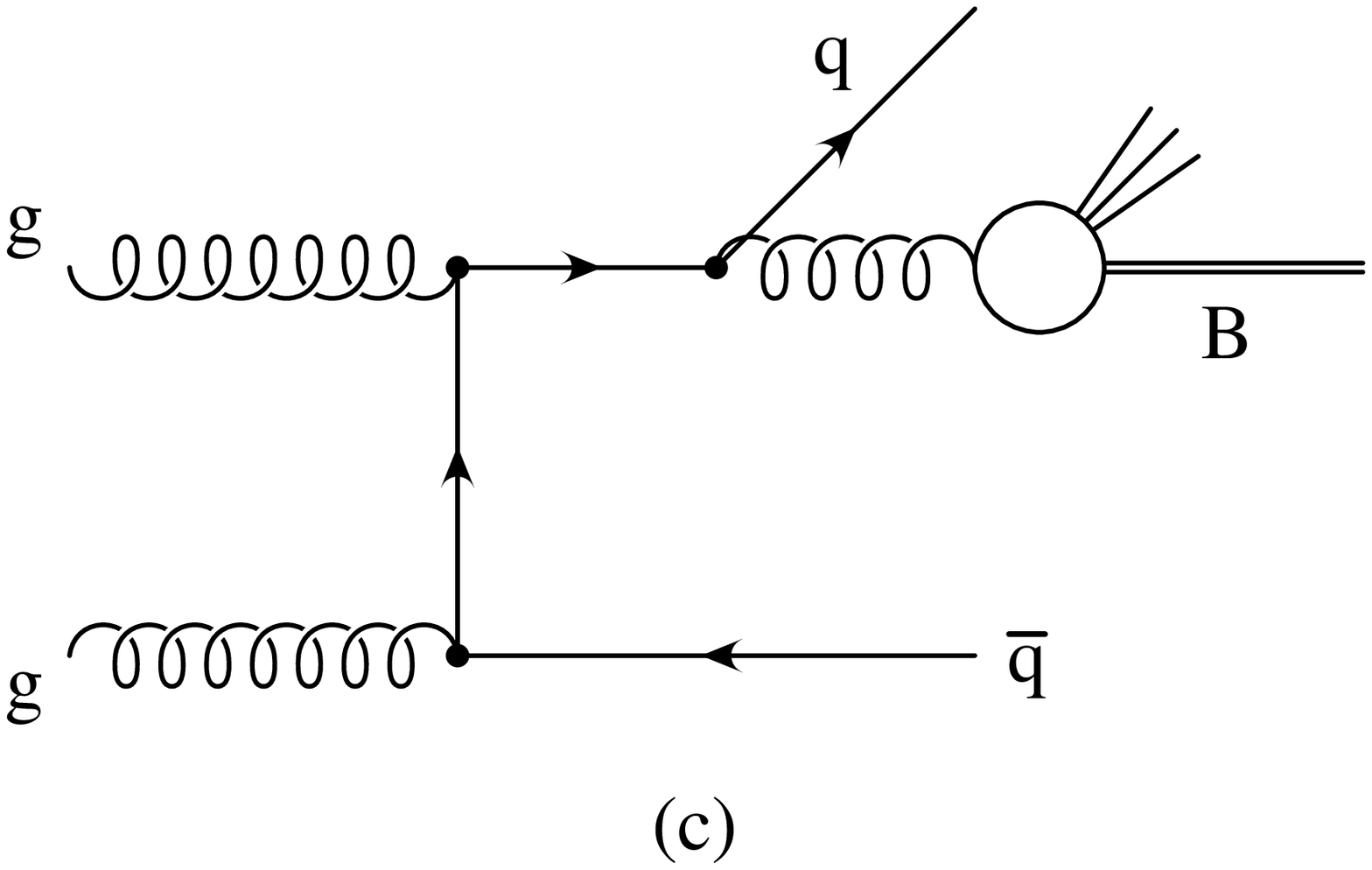,width=0.45\textwidth}}
\end{center}
\caption{Examples of Feynman diagrams leading to contributions of (a) class
(i), (b) class (ii), and (c) class (iii).}
\label{fig2}
\end{center}
\end{figure}
The calculation of the cross section $d^2\sigma/(dp_Tdy)$ of $B$-meson
hadroproduction at NLO in the GM-VFNS proceeds analogously to the case of
$D$ mesons outlined in Ref.~\cite{KKSS}.
Now, $m$ denotes the mass of the $b$ quark, and the $c$ quark belongs to the
group of light quarks $q$, whose mass is put to zero.
The NLO cross section consists of three classes of contributions.
\begin{enumerate}
\item Class (i) contains all the partonic subprocesses with a $b,\bar b\to B$
transition in the final state that have only light partons ($g,q,\bar q$) in
the initial state, the possible pairings being $gg$, $gq$, $g\bar q$, and
$q\bar q$. 
A Feynman diagram representing this class is shown in Fig.~\ref{fig2}(a).
\item Class (ii) contains all the partonic subprocesses with a $b,\bar b\to B$
transition in the final state that also have $b$ or $\bar b$ quarks in the
initial state, the possible pairings being $gb$, $g\bar b$, $qb$, $q\bar b$,
$\bar qb$, $\bar q\bar b$, and $b\bar b$ [see Fig.~\ref{fig2}(b)].
\item Class (iii) contains all the partonic subprocesses with a
$g,q,\bar q\to B$ transition in the final state [see Fig.~\ref{fig2}(c)].
\end{enumerate}
In the FFNS, only the contribution of class (i) is included, but the full $m$
dependence is retained \cite{theory}.
On the other hand, in the ZM-VFNS, the contributions of all three classes are
taken into account, but they are evaluated for $m=0$ \cite{ACGG}.
In the GM-VFNS, the class-(i) contribution of the FFNS is matched 
to the $\overline{\rm MS}$ scheme, through appropriate subtractions of
would-be collinear singularities, and is then combined with the class-(ii) and
class-(iii) contributions of the ZM-VFNS;
thus, only the hard-scattering cross sections of class (i) carry explicit $m$
dependence.
Specifically, the subtractions affect initial states involving $g\to b\bar b$
splittings and final states involving $g\to b\bar b$, $b\to gb$, and
$\bar b\to g\bar b$ splittings, and they introduce logarithmic dependences on
the initial- and final-state factorization scales in the hard-scattering cross
sections of class~(i), which are compensated through NLO by the respective
factorization scale dependences of the $b$-quark PDF and the $b\to B$ FF,
respectively.
The explicit form of the subtractions may be found in Ref.~\cite{KKSS1}.
A certain part of the class-(ii) and class-(iii) contributions is due to
Feynman diagrams with internal $b$-quark lines; another one is due to diagrams
with external $b$-quark lines and contains $m$-dependent logarithms, which are
resummed.
In the FFNS, the $m$ dependence of these contributions would only enter beyond
NLO, which is reflected in the ZM-VFNS by the generic suppression of the
$b$-quark PDF relative to the gluon and $q$-quark ones and of the gluon and
$q$-quark FFs relative to the $b$-quark one.
This entitles us to omit this $m$ dependence by calculating the contributions
of classes (ii) and (iii) in the ZM-VFNS.
It turns out that $q$-quark fragmentation contributes negligibly.
However, the gluon fragmentation contribution reaches approximately 50\% at
small values of $p_T$, and its relative contribution decreases only rather
mildly towards larger values of $p_T$.

\begin{figure}[!b]
\begin{center}
\parbox{\textwidth}{\epsfig{file=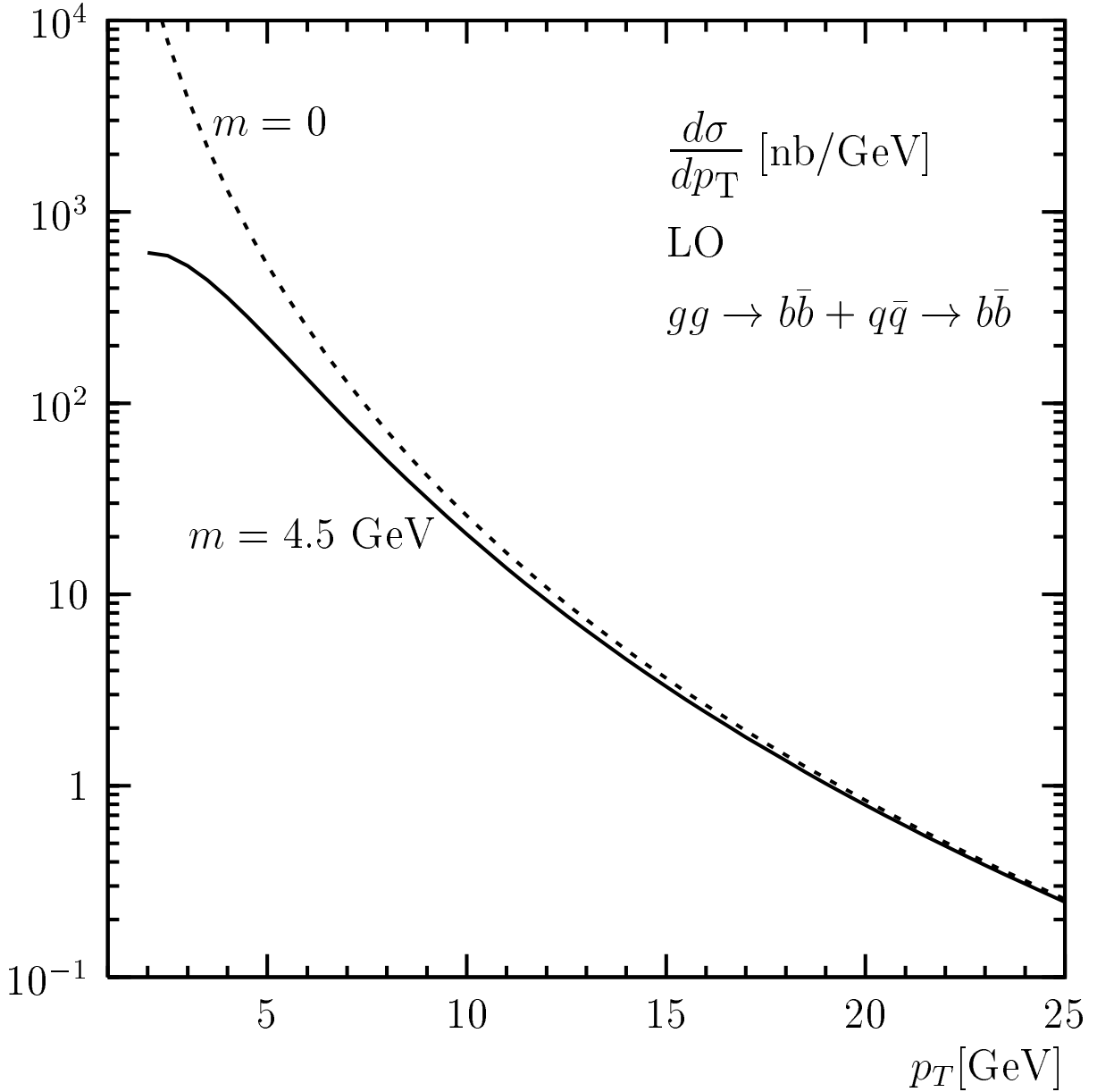,width=\textwidth}}
\caption{Transverse-momentum distribution $d\sigma/dp_T$ of
$p\bar{p} \rightarrow B +X$ at c.m.\ energy $\sqrt{S}=1.96$~TeV integrated
over the rapidity range $|y| < 1$.
The contributions of class~(i) evaluated at LO in the ZM-VFNS (dashed line)
and the GM-VFNS (solid line), but with the NLO versions of $\alpha_s$, the
PDFs, and the FFs, are compared.}
\label{fig3}
\end{center}
\end{figure}
We first investigate the effect of the finite-$m$ terms in the hard-scattering
cross sections and thus concentrate on the contribution of class (i) for the
time being.
This effect can already be studied at LO, where the partonic subprocesses read
$g + g \rightarrow b + \bar{b}$ and $q +\bar{q} \rightarrow b +\bar{b}$.
To this end, we simply switch off the NLO terms in the hard-scattering cross
sections while keeping $\alpha_s$, the PDFs, and the FFs at NLO, although,
strictly speaking, this does not represent a genuine LO analysis.
The results are shown in Fig.~\ref{fig3}, where the dashed and solid lines
refer to the results for zero and finite values of $m$, respectively.
We observe that these results rapidly approach each other with increasing
value of $p_T$.
At $p_T=7.5$~GeV, the finite-$m$ result is $33\%$ smaller than the $m=0$ one,
a relative difference of the order of $m^2/p_T^2$, as expected.

\begin{figure}[ht!] 
\begin{center}
\epsfig{file=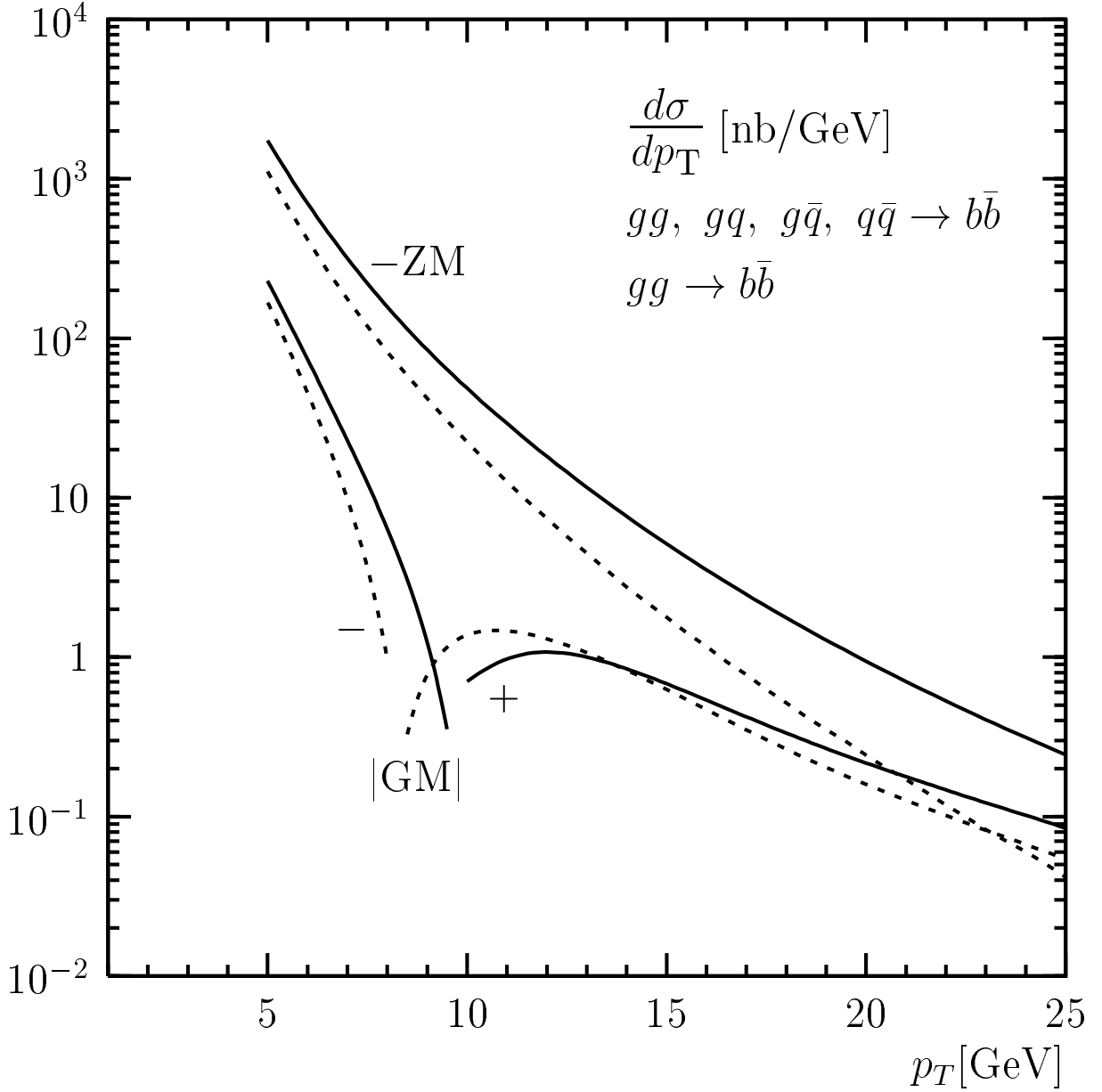,width=\textwidth}
\end{center}
\caption{Transverse-momentum distribution $d\sigma/dp_T$ of
$p\bar{p} \rightarrow B +X$ at c.m.\ energy $\sqrt{S}=1.96$~TeV integrated
over the rapidity range $|y| < 1$.
The contributions of class~(i) (solid lines) and their $gg$-initiated parts
(dashed lines) evaluated at NLO in the ZM-VFNS (upper lines) and the GM-VFNS
(lower lines) are compared.}
\label{fig4}
\end{figure}
We now turn to NLO by switching on the QCD corrections to the hard-scattering
cross sections of class~(i).
The results for $m=0$ and finite $m$ are shown in Fig.~\ref{fig4} as the upper
and lower solid lines, respectively.
They constitute parts of the final ZM-VFNS and GM-VFNS results.
In both cases, the contributions of classes (ii) and (iii) for $m=0$ still
must be added to obtain the full predictions to be compared with experimental
data.
The class-(i) contributions in the ZM-VFNS and GM-VFNS schemes are, therefore,
entitled to be negative and they indeed are, for $p_T\alt76$~GeV and
$p_T\alt10$~GeV, respectively, as may be seen from Fig.~\ref{fig4}.
Comparing the ZM-VFNS and GM-VFNS results, we notice that the finite-$m$
effects are significant for $p_T\alt10$~GeV and even cause a sign change for
10~GeV${}\alt p_T\alt76$~GeV.
However, as will become apparent below, the contributions of class (i) are
overwhelmed by those of classes (ii) and (iii), so that the finite-$m$ effects
are washed out in the final predictions, except for very small values of
$p_T$. 
It is instructive to study the relative importance of the $gg$-initiated
contributions.
They are also included in Fig.~\ref{fig4} for $m=0$ and finite $m$ as the
upper and lower dashed lines, respectively.
They exhibit a similar pattern as the full class-(i) contributions and
dominate the latter in the small-$p_T$ range.
Comparing Fig.~\ref{fig4} with Fig.~2(c) in Ref.~\cite{KKSS}, we observe that
the relative influence of the finite-$m$ effects is much smaller in the
$c$-quark case, as expected because the $c$ quark is much lighter than the $b$
quark.

\begin{figure}[bt!] 
\begin{center}
\epsfig{file=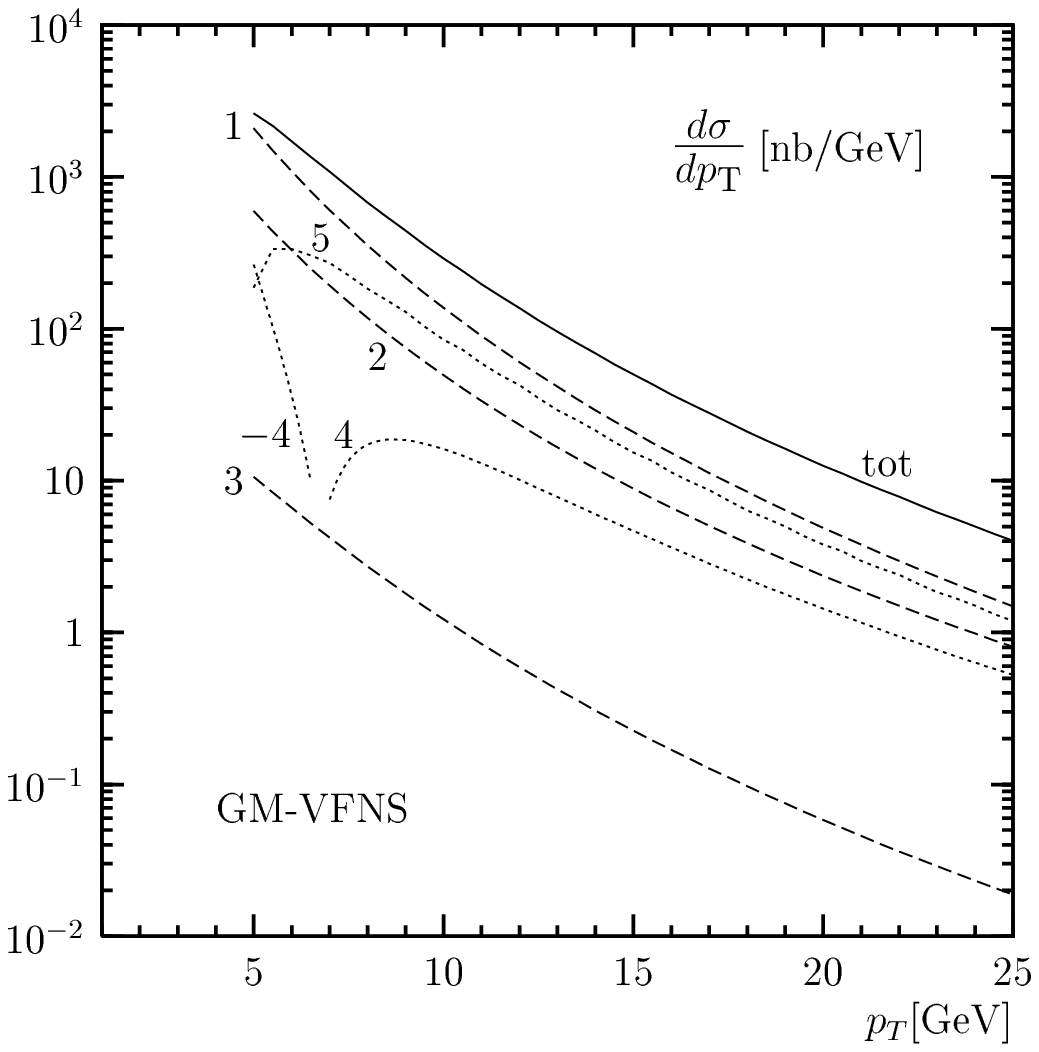,width=\textwidth}
\end{center}
\caption{Transverse-momentum distribution $d\sigma/dp_T$ of
$p\bar{p} \rightarrow B +X$ at c.m.\ energy $\sqrt{S}=1.96$~TeV integrated
over the rapidity range $|y| < 1$.
The total NLO result in the GM-VFNS including classes (i)--(iii) (solid line)
is broken up into the contributions from initial states consisting of
(1) one gluon and one $b$ (anti)quark (upper dashed line);
(2) one $q$ (anti)quark and one $b$ (anti)quark (middle dashed line);
(3) two $b$ (anti)quarks (lower dashed line);
(4) one gluon and one $q$ (anti)quark or two $q$ (anti)quarks; and
(5) two gluons.}
\label{fig5}
\end{figure}
In the remainder of this section, we work in the GM-VFNS and also include the
contributions from classes (ii) and (iii), {\it i.e.}\ we allow for $b$
(anti)quarks in the initial state and $g,q,\bar q\to B$ fragmentation.
It is interesting to study the relative importance of the various initial
states.
In Fig.~\ref{fig5}, the total result in the GM-VFNS (solid line) is
broken up into the contributions from initial states consisting of
(1) one gluon and one $b$ (anti)quark (upper dashed line);
(2) one $q$ (anti)quark and one $b$ (anti)quark (middle dashed line);
(3) two $b$ (anti)quarks (lower dashed line);
(4) one gluon and one $q$ (anti)quark or two $q$ (anti)quarks (lower dotted
line); and
(5) two gluons (upper dotted line).
If it were not for the class-(iii) contribution, then the combination of
contributions (4) and (5) would coincide with the class-(i) contribution,
considered in Fig.~\ref{fig4}, and the combination of contributions (1)--(3)
would coincide with the class-(ii) contribution.
However, in Fig.~\ref{fig5}, the class-(iii) contribution is distributed 
among the contributions (1)--(5) according to the respective initial states.
We observe from Fig.~\ref{fig5} that the partonic subprocesses with one $b$ or
$\bar b$ quark in the initial state make up the bulk of the cross section
throughout the entire mass range considered.
Specifically, the contribution from the subprocesses where the second incoming
parton is a gluon (1) is more than twice as large than the one where this is a
light (anti)quark (2), and it is even larger than the purely gluon-initiated
contribution (5), which is a surprising finding in view of the enormous gluon
luminosity in $p\bar p$ collisions at a c.m.\ energy of almost 2~TeV.
On the other hand, the contribution from two incoming $b$ (anti)quarks (3) is
greatly suppressed, being less than 1\% of the full result.
The contribution due to light-parton initial states with no more than one
gluon (4) ranks between contributions (2) and (3), and it is negative for
$p_T\alt7$~GeV.
As explained above, the difference between the $gg$-initiated GM-VFNS
contributions in Figs.~\ref{fig4} (lower dashed line) and \ref{fig5} (upper
dotted line) is due to $g,q,\bar q\to B$ fragmentation being included in the
latter.
Obviously, this additional contribution is quite significant throughout the
whole $p_T$ range considered.
Comparing the difference between the class-(i) contributions in the ZM-VFNS
and the GM-VFNS in Fig.~\ref{fig4} with the total result in Fig.~\ref{fig5},
we anticipate that the finite-$m$ effects on the latter will be rather
moderate, except for very small values of $p_T$, where the ZM-VFNS is expected
to break down anyway.
Also taking into account that the class-(i) contributions considered in
Fig.~\ref{fig4} are less negative (or even positive) in the GM-VFNS than they
are in the ZM-VFNS, we conclude that the finite-$m$ effects will moderately
enhance the cross section.
This point will be subject to further investigation in the next section.

\section{Comparison with CDF Data}
\label{sec:four}

\begin{figure}[ht!] 
\begin{center}
\epsfig{file=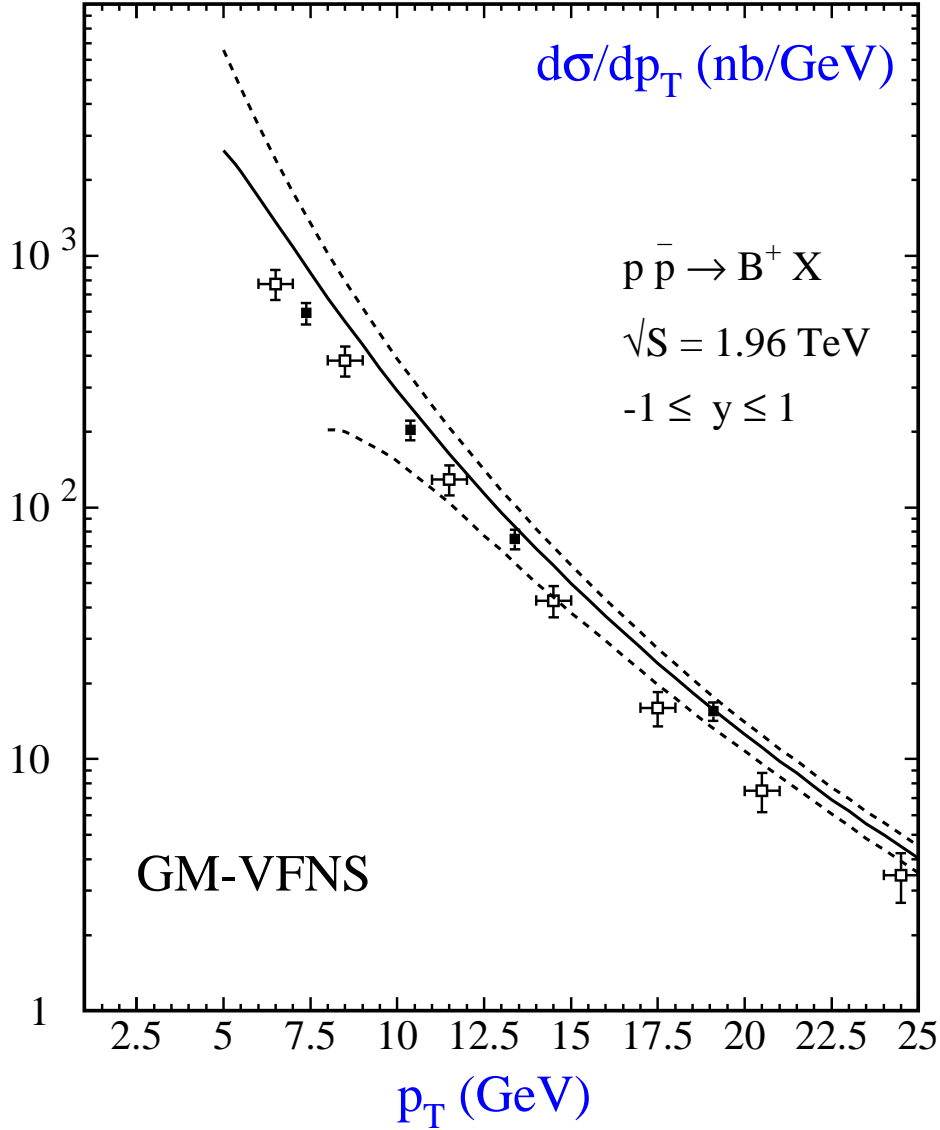,width=0.9\textwidth}
\end{center}
\caption{Transverse-momentum distribution $d\sigma/dp_T$ of
$p\bar{p} \rightarrow B +X$ at c.m.\ energy $\sqrt{S}=1.96$~TeV integrated
over the rapidity range $|y| < 1$.
The central NLO prediction with $\xi_R=\xi_F=1$ (solid line) of the GM-VFNS is
compared with CDF data from Refs.~\cite{CDF1} (open squares) and \cite{CDF2}
(solid squares).
The maximum and minimum values obtained by independently varying $\xi_R$ and
$\xi_F$ in the range $1/2\le\xi_R,\xi_F\le2$ with the constraint that
$1/2\le\xi_R/\xi_F\le2$ are also indicated (dashed lines).}
\label{fig6}
\end{figure}
We are now ready to compare our NLO predictions for the cross section
distribution $d\sigma/dp_T$ with Tevatron data.
We focus our attention on the more recent CDF data from run~II published in
Refs.~\cite{CDF1,CDF2}.
This comparison is presented for the GM-VFNS in Fig.~\ref{fig6}, where the
solid line represents the central prediction, for $\xi_R=\xi_F=1$, and the
dashed lines indicate the maximum and minimum values obtained by independently
varying $\xi_R$ and $\xi_F$ in the range $1/2\le\xi_R,\xi_F\le2$ with the
constraint that $1/2\le\xi_R/\xi_F\le2$.
The maximum and minimum values correspond to $\xi_F=2$ and $\xi_F=1/2$,
respectively.
The variation with $\xi_R$ is considerably milder than the one with $\xi_F$
and only leads to a modest broadening of the error band.
For $\xi_F<1$, $\mu_F$ reaches the starting scale $\mu_0=m$ for the DGLAP
evolution of the FFs and the $b$-quark PDF at $p_T=m\sqrt{1/\xi_F^2-1}$.
For smaller values of $p_T$, there is no prediction because the FFs and the
$b$-quark PDF are put to zero for $\mu_F<\mu_0$.
This explains why the $p_T$ distribution for $\xi_F=1/2$ only starts at
$p_T=\sqrt{3}m\approx7.8$~GeV.
The most recent data \cite{CDF2} nicely agree with the GM-VFNS result.
In fact, they lie close to the central prediction, with a tendency to fall
below it in the lower $p_T$ range, and they are comfortably contained within
the theoretical error band.
Obviously, the notorious Tevatron $B$-meson anomaly, with data-to-theory
ratios of typically 2--3 \cite{CDF}, that has been with us for more than a
decade has finally come to its end, thanks to both experimental and
theoretical progress (see also Ref.~\cite{Happacher} for a recent status
report on the observed and predicted cross sections at the Tevatron).
The previous CDF data \cite{CDF1}, based on a measurement of $J/\psi+X$ final
states, are compatible with the latest ones for $p_T\alt12$~GeV, but
systematically undershoot them for larger values of $p_T$.
This potential inconsistency becomes even more apparent by noticing that
Fig.~\ref{fig6} only contains 4 out of the 13 data points for $p_T>12$~GeV
quoted in Ref.~\cite{CDF1} and that the omitted data points neatly line up
with the selected ones.
This possibly suggests that the systematical errors in Ref.~\cite{CDF1}, and
perhaps also in Ref.~\cite{CDF2}, might be underestimated and that the overall
normalization might need some adjustment.
Incidentally, the preliminary CDF data \cite{CDF3} fall right in the middle
between those from Refs.~\cite{CDF1} and \cite{CDF2}.

The measured $p_T$ distribution of Ref.~\cite{CDF1} reaches down to $p_T=0$
and exhibits a maximum at $p_T\approx2.5$~GeV.
As we shall see below, this small-$p_T$ behavior is correctly reproduced in
the FFNS without DGLAP-evolved FFs, which only receives contributions of
class (i) without any subtractions.
It is clear that our present implementation of the GM-VFNS is not suitable for
cross section calculations in the small-$p_T$ region.
Although the GM-VFNS is designed to approach the FFNS and the ZM-VFNS in its
regions of validity without introducing additional ad-hoc matching factors,
to implement this numerically is a non-trivial task due to necessary
cancellations between different terms in the calculation.
Stable computer codes including these features have been developed for the
fully inclusive case and are used in global analyses of proton PDFs,
{\it e.g.}\ in the CTEQ studies \cite{CTEQ} using the ACOT scheme \cite{ACOT}.
For one-particle inclusive processes, the problem to achieve such
cancellations is complicated by the extra factorization scale; to obtain a
smooth transition from the GM-VFNS to the FFNS, one has to carefully match
terms that are taken into account at fixed order with terms that are resummed
to higher orders in the PDFs and FFs.
In addition, it remains to be investigated whether a proper scale choice in
the small-$p_T$ range is required and helpful to ensure that the FFs and
$b$-quark PDF are sufficiently suppressed already at $p_T={\cal O}(m)$.

The GM-VFNS prediction in Fig.~\ref{fig6} exhibits a sizeable scale
uncertainty for $p_T\alt2m$.
As mentioned above, the $p_T$ distribution for $\xi_F=1/2$ only starts at
$p_T\approx7.8$~GeV.
These undesirable features will eventually be removed once the matching with
the FFNS is specified and implemented.
This is a so-called implementation issue \cite{Tung:2001mv} that needs to be
added on top of the definition of the pure GM-VFNS.
This is beyond the scope of the present paper, which is concerned with the
intermediate $p_T$ range, and will be treated in a future publication.
At this point, we would like to recall how this implementation issue is
handled for the FONLL scheme \cite{CGN}.
In that scheme, the FFNS and ZM-VFNS calculations are merged in such a way
that the contribution that is added on to the FFNS result, {\it i.e.}\ the
ZM-VFNS result with the zero-mass limit of the FFNS result subtracted, is
multiplied by a weight function of the form $p_T^2/(p_T^2+c^2m^2)$ with $c=5$
to model a smooth transition.
Furthermore, the variable $p_T$ of the subtracted ZM-VFNS contribution is
shifted to become $m_T$.
In the region where the ZM-VFNS prediction has a large scale uncertainty,
{\it i.e.}\ where $p_T$ is 2--3 times larger than $m$ say, this weight
function is still rather small, ranging from 14\% to 26\%.
Thus, this weight function not only smoothens the transition, but also ensures
that the sizeable theoretical uncertainty of the ZM-VFNS component in the
transition region is not reflected in the FONLL prediction, creating the
impression that the latter has a small theoretical error.
Of course, this source of theoretical uncertainty unavoidably resurfaces when
the form of the weight function, which is a priori unknown, is varied,
{\it e.g.}\ by changing the value of its parameter $c$.
Unfortunately, such a variation is not included in the theoretical error of
recent FONLL predictions \cite{CN,Cacciari:2003uh}.

\begin{figure}[!thb]
\begin{center}
\epsfig{file=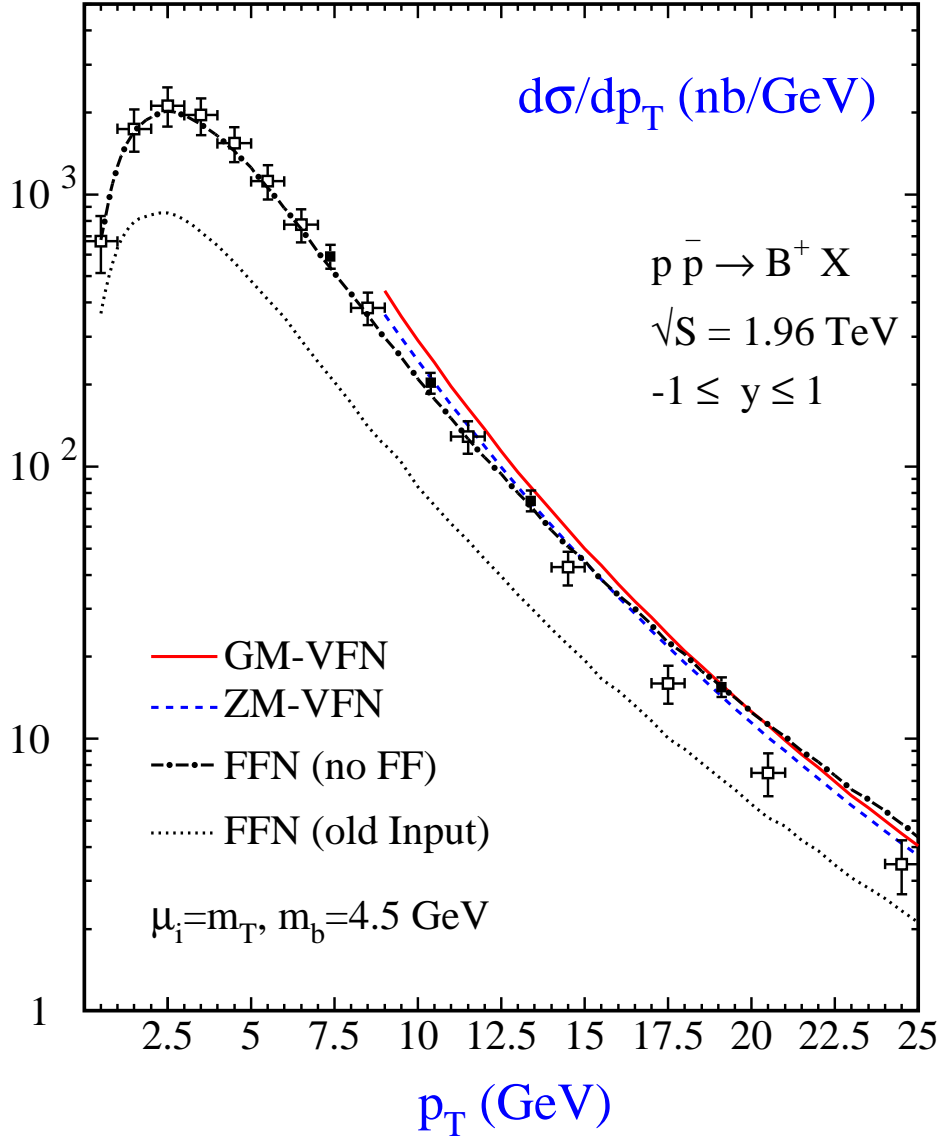,width=0.9\textwidth}
\caption{Transverse-momentum distribution $d\sigma/dp_T$ of
$p\bar{p} \rightarrow B +X$ at c.m.\ energy $\sqrt{S}=1.96$~TeV integrated
over the rapidity range $|y| < 1$.
The central NLO predictions in the FFNS with $n_f=4$ and without FFs
(dot-dashed line), the ZM-VFNS (dashed line), and the GM-VFNS (solid line) are
compared with CDF data from Refs.~\cite{CDF1} (open squares) and \cite{CDF2}
(solid squares).
For reference, the historical FFNS prediction, evaluated with PDF set MRSD0
\cite{Martin:1992as}, a $b\to B$ FF of Peterson type \cite{Peterson} with
$\epsilon=0.006$, $m_b=4.75$~GeV, and
$\Lambda_{\overline{\rm MS}}^{(4)}=215$~MeV, is also shown.}
\label{fig7}
\end{center}
\end{figure}
We now extend our numerical analysis to include the NLO prediction in the
FFNS, with $n_f=4$ massless quark flavors in the initial state, which allows
us to also compare with the small-$p_T$ data from Ref.~\cite{CDF1}.
In the FFNS analysis, we evaluate $\alpha_s^{(n_f)}(\mu_R)$ with $n_f=4$ and
$\Lambda_{\overline{\rm MS}}^{(4)} = 326$~MeV \cite{CTEQ}, while we continue
using the CTEQ6.1M proton PDFs \cite{CTEQ}, in want of a rigorous FFNS set
with $n_f=4$.
In the FFNS, there is no room for DGLAP-evolved FFs, and only $b,\bar b\to B$
transitions are included.
For simplicity, we identify $b$ (anti)quarks with $B$ mesons and account for
non-perturbative effects by including the branching fraction
$B(b\to B)=39.8\%$ \cite{pdg} as an overall normalization factor, {\it i.e.}\
we use a $b\to B$ FF of the form $D(x)=B(b\to B)\delta(1-x)$, while the
$g,q,\bar q\to B$ FFs are put to zero.
In Fig.~\ref{fig7}, the central FFNS (dot-dashed line), ZM-VFNS (dashed line),
and GM-VFNS (solid line) predictions, for $\xi_R=\xi_F=1$, are compared with
the CDF data from Refs.~\cite{CDF1,CDF2}.
As in Fig.~\ref{fig6}, some of the data points with $p_T>7$~GeV from
Ref.~\cite{CDF1} are omitted for clarity.
Since the ZM-VFNS and our present implementation of the GM-VFNS are not
applicable to the small-$p_T$ range, we show the respective predictions only
for $p_T>2m=9$~GeV.
The GM-VFNS prediction shown in Fig.~\ref{fig7} is identical with the central
one in Fig.~\ref{fig6}.
By construction, it merges with the ZM-VFNS prediction with increasing value
of $p_T$.
In accordance with the expectation expressed in the discussion of
Figs.~\ref{fig3} and \ref{fig4}, the difference between the GM-VFNS and
ZM-VFNS results is rather modest also at $p_T\agt2m$, since the $m$-dependent
contribution, of class~(i), is numerically small and overwhelmed by the
$m$-independent ones, of classes (ii) and (iii).
The FFNS prediction faithfully describes the peak structure exhibited by the
next-to-latest CDF data \cite{CDF1} in the small-$p_T$ range and it also
nicely agrees with the latest CDF data \cite{CDF2} way out to the largest
$p_T$ values.
In fact, for $p_T>4m$, where its perturbative stability is jeopardized by
unresummed logarithms of the form $\ln(m_T^2/m^2)\agt3$, the FFNS prediction
almost coincides with the GM-VFNS one, where such large logarithms are
resummed.
This is a pure coincidence, which becomes even more apparent if we also recall
that the implementation of the $b,\bar b\to B$ transition in the FFNS is not
based on a factorization theorem and quite inappropriate for such large values
of $p_T$.

\begin{figure}[!thb]
\begin{center}
\epsfig{file=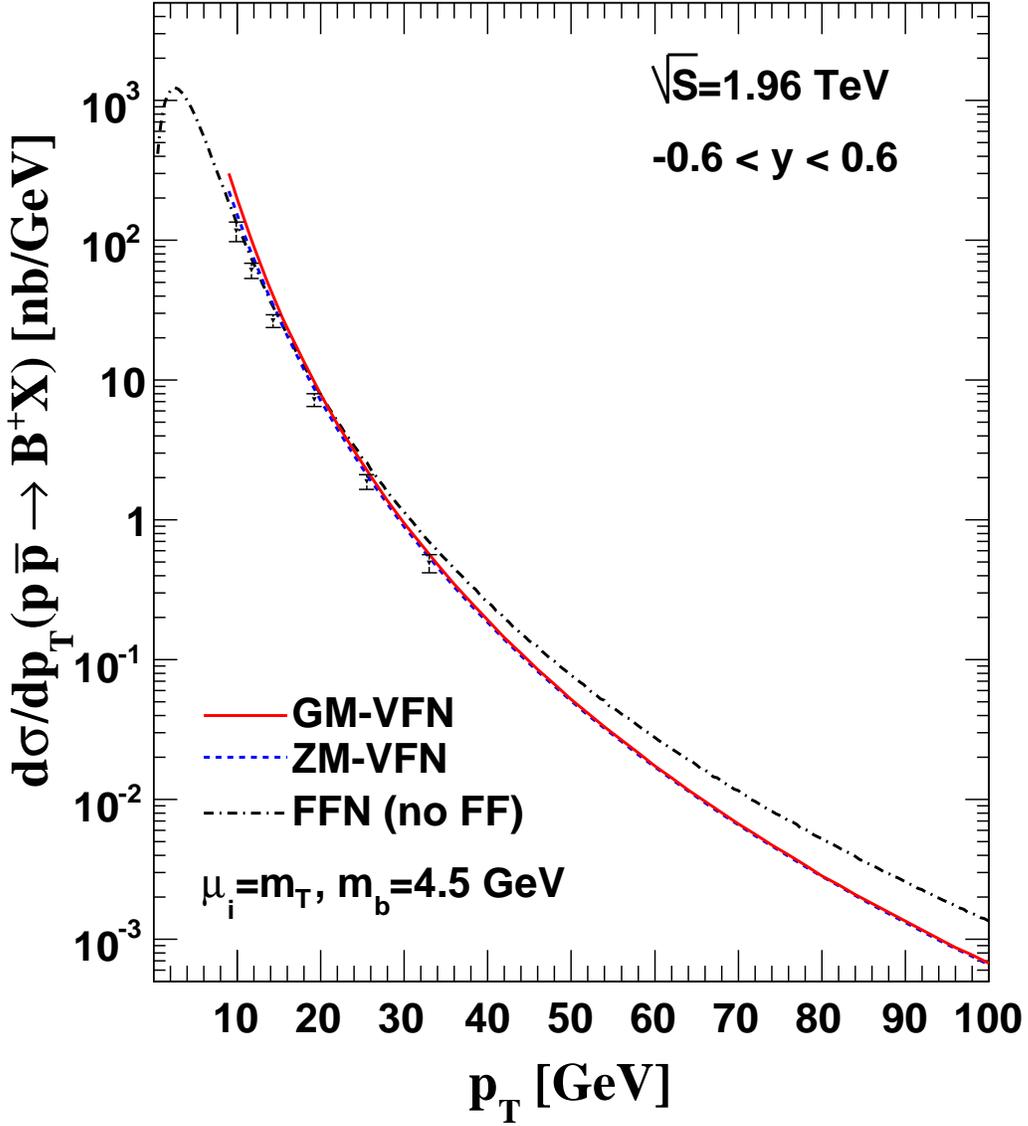,width=0.9\textwidth}
\caption{Transverse-momentum distribution $d\sigma/dp_T$ of $p\bar{p}\to B+X$
at c.m.\ energy $\sqrt{S}=1.96$~TeV integrated over the rapidity range
$|y|<0.6$.
The central NLO predictions in the FFNS with $n_f=4$ and without FFs
(dot-dashed line), the ZM-VFNS (dashed line), and the GM-VFNS (solid line) are
compared in the large-$p_T$ range.}
\label{fig8}
\end{center}
\end{figure}
In Figs.~\ref{fig6} and \ref{fig7}, we limited our considerations to the range
$p_T<25$~GeV, where the published CDF data \cite{CDF1,CDF2} are located.
However, the preliminary CDF data \cite{CDF3}, collected in the very central
part of the detector ($|y|<0.6$), cover the range 9~GeV${}<p_T<40$~GeV, and it
is interesting to confront them with the NLO predictions of the three schemes
considered here.
Moreover, it is instructive to study the breakdown of the FFNS at sufficiently
large values of $p_T$ due to unresummed large logarithms.
For these purposes, we show in Fig.~\ref{fig8} an extension of Fig.~\ref{fig7}
including the preliminary CDF data.
We observe that the GM-VFNS result steadily merges with the ZM-VFNS one as the
value of $p_T$ is increased, the relative deviations being 11\%, 6\%, and 3\%
at $p_T=20$, 30, and 50~GeV, respectively.
The FFNS result breaks even with the GM-VFNS one at about $p_T=20$~GeV (see
also Fig.~\ref{fig7}) and exceeds the latter for larger values of $p_T$, the
relative deviations being 20\%, 48\%, and 100\% at $p_T=30$, 50, and 100~GeV,
respectively.
Our results indicate that a measurement of the $p_T$ distribution up to
40--50~GeV at the Tevatron could be able to resolve the difference between the
FFNS and the VFNS and so to establish for the first time the break-down of the
FFNS due to unresummed logarithms in the inclusive hadroproduction of heavy
hadrons.
This important question deserves a careful examination of the theoretical
uncertainties, which we leave for a future publication. 
The preliminary CDF data point in the bin 29~GeV${}<p_T<40$~GeV favors the
ZM-VFNS and GM-VFNS results, while it undershoots the FFNS result.

We conclude this section with an interesting observation that, in retrospect,
sheds some new light on the Tevatron $B$-hadron anomaly mentioned above and
does not appear to be sufficiently well known to the community.
In fact, the common perception that the CDF data \cite{CDF,CDF1,CDF2}
generally overshoot the FFNS prediction, frequently denoted as NLO QCD in the
literature, by a factor of 2--3 is entirely due the use of obsolete
theoretical input.
In fact, the FFNS prediction that has been serving as a benchmark for some 15
years and still does even in very recent papers \cite{CDF2,Happacher} is
evaluated with the proton PDF set MRSD0 by Martin, Roberts, and Stirling
\cite{Martin:1992as}, which has been revoked by these authors.
It has an unacceptably weak gluon and a small value of
$\Lambda_{\overline{\rm MS}}^{(4)}$, namely
$\Lambda_{\overline{\rm MS}}^{(4)}=215$~MeV translating into
$\alpha_s^{(5)}(m_z)=0.111$, which is 3.3 standard deviations below the
present world average $\alpha_s^{(5)}(m_z)=0.1176\pm0.0020$ \cite{pdg}.
Other inputs include $m_b=4.75$~GeV and a Peterson FF parameter of
$\epsilon=0.006$, extracted from a fit to $e^+e^-$ annihilation data from the
pre-LEP/SLC era using a Monte-Carlo event generator based on massless LO
matrix elements \cite{Chrin:1987yd}.
For reference, the historical FFNS prediction evaluated with this choice of
input is also included in Fig.~\ref{fig7}.
Since the FONLL prediction \cite{CN} is designed to merge with the FFNS one at
low values of $p_T$, the additional contribution being faded out by a weight
function of the form $p_T^2/(p_T^2+25m_b^2)$, the striking gap between the
historical NLO QCD prediction and the FONLL prediction, based on up-to-date
input information, in Fig.~10 of Ref.~\cite{CDF2} impressively illustrates the
advancement in the PDF and $\alpha_s$ determinations.
The tuning of FFs in connection with the resummation of leading and
next-to-leading logarithms, emphasized in the second paper of
Ref.~\cite{CDF3}, is actually of minor importance.

\section{Conclusions}
\label{sec:five}

For several years, the $B$-meson production rates measured at DESY HERA, CERN
LEP2, and Tevatron have been notoriously exceeding, by up to a factor of
three, the usual NLO QCD predictions for massive $b$ quarks, {\it i.e.}\ those
in the FFNS ($B$-meson anomaly).
This has even triggered theoretical attempts to interpret this deviation as a
signal of new physics beyond the standard model \cite{Berger:2000mp}.
However, it remained to be clarified if this deviation could be explained by
improving and refining the QCD prediction itself.
In this connection, two of us, together with Binnewies, pointed out almost a
decade ago that the ZM-VFNS provides a rigorous theoretical framework for a
coherent study of $B$-meson production in high-energy $e^+e^-$, $p\bar p$, and
other collisions, since the factorization theorem guarantees the universality
of the $B$-meson FFs \cite{BKK}.
In fact, the ZM-VFNS prediction \cite{BKK,BK} was found to nicely agree with
the CDF data from Tevatron runs IA and I \cite{CDF}.
However, a necessary condition for the applicability of the ZM-VFNS is that
the energy scale that separates perturbative hard scattering from
non-perturbative fragmentation (final-state factorization scale $\mu_F$) is
sufficiently large compared to the $b$-quark mass $m$, and it had never been
quite clear how large the ratio $\mu_F/m$ actually needed to be in order for
finite-$m$ effects to be negligible.
In fact, the authors of Ref.~\cite{Cacciari:2003uh} asserted in a footnote that
mass corrections have a large size up to $p_T\approx20$~GeV and that ``lack of
mass effects \cite{BKK} will therefore erroneously overestimate the production
rate at small $p_T$.''

In the present paper, we addressed this problem by performing a comparative
analysis of $B$-meson hadroproduction in the ZM-VFNS and the GM-VFNS, which we
had successfully applied to $D$-meson production in $\gamma\gamma$ \cite{KS},
$ep$ \cite{KS1}, and $p\bar p$ \cite{KKSS,KKSS1,KKSS2} collisions in the past.
For this, we also updated the determination of $B$-meson FFs \cite{BKK} by
fitting to recent $e^+e^-$ data from ALEPH \cite{Aleph}, OPAL \cite{Opal1},
and SLD \cite{SLD} and also adjusting the values of $m$ and the energy scale
$\mu_0$ where the DGLAP evolution starts to conform with modern PDF sets
\cite{CTEQ}.
We found that finite-$m$ effects moderately enhance the $p_T$ distribution;
the enhancement amounts to about 20\% at $p_T=2m$ and rapidly decreases with
increasing value of $p_T$, falling below 10\% at $p_T=4m$.
This finding contradicts earlier claims~\cite{Cacciari:2003uh} in all respects.
Such effects are comparable in size to the theoretical uncertainty due to the
freedom of choice in the setting of the renormalization and factorization
scales.
For comparison, we also evaluated the $p_T$ distribution in the FFNS, with
$n_f=4$, using a delta-function-type $b\to B$ FF without DGLAP evolution.

Confronting the three NLO predictions with the latest \cite{CDF2} and
next-to-latest \cite{CDF1} CDF data sets published, we found that all of them
agree rather well with the latest one, with $p_T>7$~GeV.
Despite unresummed large logarithms and poorly implemented fragmentation, the
FFNS prediction happens to almost coincide with the GM-VFNS one in the range
15~GeV${}\alt p_T\alt25$~GeV.
The FFNS prediction also nicely reproduces the peak exhibited about
$p_T\approx2.5$~GeV by the next-to-latest CDF data \cite{CDF1}.
By contrast, the historical benchmark result based on obsolete proton PDFs and
a value of $\alpha_s^{(5)}(m_z)$ falling short of the present world average by
3.3 standard deviations, which goes under the name NLO QCD in the literature
and is used as a reference point even in most recent papers
\cite{CDF2,Happacher}, undershoots the CDF data by the familiar factor of
2--3.
This illustrates that the progress in our understanding of the proton PDFs and
our knowledge of $\alpha_s^{(5)}(m_Z)$ is instrumental in overcoming the
long-standing Tevatron $B$-hadron anomaly in the low to intermediate $p_T$
range.
The preliminary CDF data \cite{CDF3} favor the ZM-VFNS and GM-VFNS results in
the upmost bin, 29~GeV${}<p_T<40$~GeV, while they undershoot the FFNS result.

It is desirable to extend the applicability of the GM-VFNS down to $p_T=0$.
This requires matching with the FFNS.
To achieve this in a way that avoids ad-hoc weight functions is a non-trivial
task and is left for future work.

\section*{Acknowledgement}

The work of BAK and GK was supported in part by the German Federal Ministry
for Education and Research BMBF through Grant No.\ 05~HT6GUA and by the German
Research Foundation DFG through Grant No.\ KN~365/7--1.

\end{document}